\documentclass[prc,showpacs,superscriptaddress,twocolumn]{revtex4}

\usepackage{amsmath,bm}
\usepackage{graphicx}
\usepackage{epstopdf}
\usepackage{subfigure}
\usepackage{epsfig}
\usepackage{amsmath,amssymb,amsfonts}
\usepackage{color}
\usepackage[utf8]{inputenc}
\usepackage{verbatim}
\usepackage{array}
\graphicspath{{figures/}} 

\begin{document}

\newcommand{\vk}{{\vec k}}
\newcommand{\vK}{{\vec K}}
\newcommand{\vb}{{\vec b}}
\newcommand{{\vp}}{{\vec p}}
\newcommand{{\vq}}{{\vec q}}
\newcommand{\vQ}{{\vec Q}}
\newcommand{\vx}{{\vec x}}
\newcommand{\beq}{\begin{equation}}
\newcommand{\eeq}{\end{equation}}
\newcommand{\half}{{\textstyle \frac{1}{2}}}
\newcommand{\gton}{\stackrel{>}{\sim}}
\newcommand{\lton}{\mathrel{\lower.9ex \hbox{$\stackrel{\displaystyle<}{\sim}$}}}
\newcommand{\ee}{\end{equation}}
\newcommand{\ben}{\begin{enumerate}}
\newcommand{\een}{\end{enumerate}}
\newcommand{\bit}{\begin{itemize}}
\newcommand{\eit}{\end{itemize}}
\newcommand{\bc}{\begin{center}}
\newcommand{\ec}{\end{center}}
\newcommand{\bea}{\begin{eqnarray}}
\newcommand{\eea}{\end{eqnarray}}

\newcommand{\beqar}{\begin{eqnarray}}
\newcommand{\eeqar}[1]{\label{#1} \end{eqnarray}}
\newcommand{\pleft}{\stackrel{\leftarrow}{\partial}}
\newcommand{\pright}{\stackrel{\rightarrow}{\partial}}

\newcommand{\eq}[1]{Eq.~(\ref{#1})}
\newcommand{\fig}[1]{Fig.~\ref{#1}}
\newcommand{\eff}{ef\!f}
\newcommand{\alphas}{\alpha_s}

\renewcommand{\topfraction}{0.85}
\renewcommand{\textfraction}{0.1}
\renewcommand{\floatpagefraction}{0.75}

\title{Radial profile of bottom quarks in jets in high-energy nuclear collisions}

\author{Sa Wang}
\affiliation{Key Laboratory of Quark \& Lepton Physics (MOE) and Institute of Particle Physics,
 Central China Normal University, Wuhan 430079, China}
\affiliation{Guangdong Provincial Key Laboratory of Nuclear Science, Institute of Quantum Matter, South China Normal University, Guangzhou 510006, China}
\affiliation{Guangdong-Hong Kong Joint Laboratory of Quantum Matter, Southern Nuclear Science Computing Center, South China Normal University, Guangzhou 510006, China}

\author{Wei Dai}
\affiliation{School of Mathematics and Physics, China University
of Geosciences, Wuhan 430074, China}

\author{Ben-Wei Zhang}
\email{bwzhang@mail.ccnu.edu.cn}
\affiliation{Key Laboratory of Quark \& Lepton Physics (MOE) and Institute of Particle Physics, Central China Normal University, Wuhan 430079, China}
\affiliation{Guangdong Provincial Key Laboratory of Nuclear Science, Institute of Quantum Matter, South China Normal University, Guangzhou 510006, China}
\affiliation{Guangdong-Hong Kong Joint Laboratory of Quantum Matter, Southern Nuclear Science Computing Center, South China Normal University, Guangzhou 510006, China}

\author{Enke Wang}
\affiliation{Guangdong Provincial Key Laboratory of Nuclear Science, Institute of Quantum Matter, South China Normal University, Guangzhou 510006, China}
\affiliation{Guangdong-Hong Kong Joint Laboratory of Quantum Matter, Southern Nuclear Science Computing Center, South China Normal University, Guangzhou 510006, China}
\affiliation{Key Laboratory of Quark \& Lepton Physics (MOE) and Institute of Particle Physics,
 Central China Normal University, Wuhan 430079, China}

\date{\today}


\begin{abstract}
Angular correlations between heavy quark~(HQ) and its tagged jet are potentially new tools to gain insight into the in-medium partonic interactions in relativistic heavy-ion collisions. In this work, we present the first theoretical study on the radial profiles of B mesons in jets in Pb+Pb collisions at the LHC.  The initial production of bottom quark tagged jet in p+p is computed by SHERPA which matches the next-to-leading order matrix elements with contributions of parton shower, whereas the massive quark traversing the QGP described by a Monte Carlo model SHELL which can simultaneously simulate light and heavy flavor in-medium energy loss within the framework of Langevin evolution. In p+p collisions, we find that at lower $\rm p_T^Q$ the radial profiles of heavy flavors in jets are sensitive to the heavy quark mass. In $0-10\%$ Pb+Pb collisions at $\rm \sqrt{s_{NN}}=5.02$~TeV, we observe an inverse modification pattern of the B mesons radial profiles in jets at $\rm 4<p_T^Q<20$~GeV compared to that of D mesons: the jet quenching effects narrow the jet radial profile of B mesons in jets while broaden that of D mesons in jets. We find that in A+A collisions, the contribution dissipated from the higher $\rm p_T^Q> 20$~GeV region naturally has a narrower initial distribution and consequently leads to a narrower modification pattern of radial profile; however the diffusion nature of the heavy flavor in-medium interactions will give rise to a broader modification pattern of radial profile. These two effects consequently compete and offset with each other, and the b quarks in jets benefit more from the former and suffers less diffusion effect compared to that of c quarks in jets. These findings can be tested in the future experimental measurements at the LHC to gain better understanding of the mass effect of jet quenching.

\end{abstract}


\maketitle

\section{Introduction}
\label{sec:intro}
High-energy nuclear collisions at the Relativistic Heavy Ion Collider (RHIC) and the Large Hadron Collider (LHC) provide excellent arena to unravel the properties of the quark-gluon plasma (QGP), a new state of nuclear matter with de-confined quarks and gluons, which is predicted to be formed at extreme hot and dense system by the Quantum Chromodynamics (QCD), the fundamental theory of strong interaction.
In the past few decades, the ``jet quenching" phenomenon, the energy loss of the initial-produced energic jet due to the strong interactions with the the constituents of the QGP,  aroused  physicists' great interest and has been extensively studied ~\cite{Gyulassy:2003mc,Qin:2015srf,Vitev:2008rz,Vitev:2009rd,CasalderreySolana:2010eh,He:2011pd,Neufeld:2010fj,Senzel:2013dta,Casalderrey-Solana:2014bpa,Dai:2012am,Zhang:2018urd,Milhano:2015mng,Chang:2016gjp,Connors:2017ptx,
Zhang:2018ydt,Cao:2020wlm,Yan:2020zrz,Chen:2020pfa}. These studies show that the differences between the final-state observables at large transverse momentum, such as leading hadron spectra and jet production, in p+p and A+A collisions can help us gain insight into the mechanisms of in-medium parton interactions and precisely extract valuable information on the properties of the QGP created in relativistic heavy-ion collisions.

Owing to the large mass ($M_Q\gg T$) and early creation time, heavy quarks are witnesses of the entire QGP evolution, therefore viewed as ideal hard probes to constrain the transport properties of QGP and also improve our understanding of the in-medium heavy quarks evolution. As new favorites of observables in heavy-ion collisions (HIC), the nuclear modification factor $R_{AA}$~\cite{Adamczyk:2014uip,Adam:2015sza,Sirunyan:2017xss}, collective flow $v_n$~\cite{Abelev:2014ipa,Adamczyk:2017xur,Acharya:2017qps,Sirunyan:2017plt} of heavy flavored mesons and $D^0$+hadron correlations~\cite{Adam:2019tql} are extensively measured in experiment and successfully modeled in theory~\cite{ vanHees:2007me,CaronHuot:2008uh,vanHees:2005wb,Djordjevic:2015hra,He:2014cla,Chien:2015vja,Kang:2016ofv,Cao:2013ita,Alberico:2013bza,Xu:2015bbz,Cao:2016gvr,
Das:2016cwd,Ke:2018tsh,Xing:2019xae,Li:2020kax,Li:2020umn,Altenkort:2020fgs,Jamal:2020fxo,He:2019vgs}, however, there are still some important key questions to be addressed~\cite{Dong:2019unq}, for reviews see Refs.~\cite{Andronic:2015wma,Cao:2018ews,Dong:2019unq,Dong:2019byy,Cao:2021ces,Zhao:2020jqu}.

The recent measurements relating to the production and substructure of heavy flavor jets shed new light on the jet physics in high-energy heavy-ion collisions~\cite{Chatrchyan:2013exa,Sirunyan:2018jju,Sirunyan:2019dow}, aiming to address the mass effect on in-medium jet shower evolution. Among them, the nuclear modification factor $R_{pA}$~\cite{Khachatryan:2015sva,Sirunyan:2016fcs} and $R_{AA}$~\cite{Chatrchyan:2013exa} of heavy flavor jet offer strong tools for quantifying the cold nuclear matter (CNM) effects in initial-state and the in-medium quenching effects in the subsequent formed QGP~\cite{Li:2017wwc,Li:2018xuv,Kang:2018wrs,Dai:2018mhw,Wang:2019xey}. Especially, the recent reported radial distributions of charm meson in jets by CMS collaboration provide an interesting opportunity to investigate the diffusions effects of heavy quarks as well as the modified substructure of heavy flavor jets in nucleus-nucleus collisions from a new angle~\cite{Sirunyan:2019dow}, and the detailed discussion for c-jet has been presented in our previous work~\cite{Wang:2019xey}. Since bottom quarks have larger mass than charm quark~($m_{b}\sim 4.5$~GeV, $m_{c}\sim 1.5$~GeV), the radial profile of bottom quarks in jets can be new signal to gain insight into the in-medium partonic interaction and the mass effects. Therefore, it will be interesting and urgent to investigate the radial distribution of bottom quarks in jets in p+p collisions and its modification in nucleus-nucleus collisions.

In this work, we present the first theoretical study on the radial distribution of bottom quarks in jets in heavy-ion collisions. A systematic comparison between the radial profile of bottom and charm quarks in jets are also investigated, aiming to figure out the sensitivity of quark mass in the heavy quark diffusion effects. We employ a Monte Carlo event generator SHERPA~\cite{Gleisberg:2008ta} at NLO+PS~(next-to-leading order QCD calculations matched with parton shower) accuracy as p+p baseline, and also take into account the in-medium elastic and inelastic parton energy loss. In p+p collisions, we find that the radial profiles of heavy quarks in jets are sensitive to the heavy quark mass. In $0-10\%$ Pb+Pb collisions at $\sqrt{s_{NN}}=5.02$~TeV, we observe an inverse modification pattern of the radial profiles of bottom quarks in jets compared to that of charm quarks: jet quenching effects will narrow the jet radial profile of bottom quarks while broaden that of charm quarks. And we will demonstrate that the different modification patterns are mainly determined by their initial radial distributions in jets.

The remainder of this paper is organized as follows. In Sec.~II, we present the radial distributions of bottom quarks in jets in p+p collisions compared to charm quarks. In Sec.~III, the Monte Carlo framework of the in-medium jet evolution are introduced. Discussions on the medium modification of the bottom quarks radial profile in jets are given in Sec.~IV. We will summarize this paper in Sec.~V.

\section{Radial distribution of bottom quarks in jets in p+p collisions}
\label{sec:ppbaseline}

The productions of heavy flavor jets~(charm jet and bottom jet, usually denotes as c-jet and b-jet) in high energy physics are extensively studied both as tests of perturbative QCD calculation~\cite{Andersson:1983ia} and probes for other physics aspects within and beyond the Standard Model~\cite{Norrbin:2000zc}. In definition, heavy quark jets are defined as the jets containing heavy quark~(heavy flavor meson) inside the jet. In general, the production mechanisms of heavy flavor jets in hadron collisions are often attributed into three categories: flavor creation~(FCR), flavor excitation~(FEX), and gluon splitting~(GSP)~\cite{Norrbin:2000zc,Banfi:2007gu}. FCR represents the $Q\bar{Q}$ pairs creation process at leading order~($q+\bar{q}\rightarrow Q+\bar{Q}$, $g+g\rightarrow Q+\bar{Q}$) in the hard scattering. FEX is process that a heavy quark from the initial parton distribution of one beam particle is excited onto mass shell by hard scattering by a parton of the other beam particle. GSP is the situation that $g\rightarrow Q+\bar{Q}$ branching occurs in the initial or final state parton shower, but not involves the hard scattering. In this work, the heavy flavor jet productions in p+p collisions are provided by the Monte Carlo event generator SHERPA~\cite{Gleisberg:2008ta} in which the NLO QCD matrix elements are matched with parton shower (PS) availably~\cite{Frixione:2002ik}. The NNPDF3.0 ~\cite{Ball:2014uwa} NLO parton distribution function (PDF) with massive bottom quark was chosen in the calculations. The final-state jet reconstruction and event selection are implemented with anti-$k_T$ algorithm~\cite{Cacciari:2008gp} within \textsl{Fastjet} package~\cite{Cacciari:2011ma}. In our current study,  the fragmentation of heavy quarks~($c\rightarrow D$, $b\rightarrow B$) is performed by introducing the Peterson form fragmentation functions (FFs)~\cite{Peterson:1982ak} $D(z)\propto1/z(1-1/z-\epsilon/(1-z))^2$, where $\epsilon_c=0.01, \epsilon_b=0.001$~\cite{Das:2016llg,Cacciari:2005rk,Cacciari:2012ny}. According to the estimates in Ref.~\cite{Cao:2013ita}, the in-medium coalescence mechanism has less influence on the heavy quarks with $p_T>4~$GeV.



\begin{figure}[!t]
\begin{center}
\vspace*{0.1in}
\hspace*{-.1in}
\includegraphics[width=3.4in,height=5.1in,angle=0]{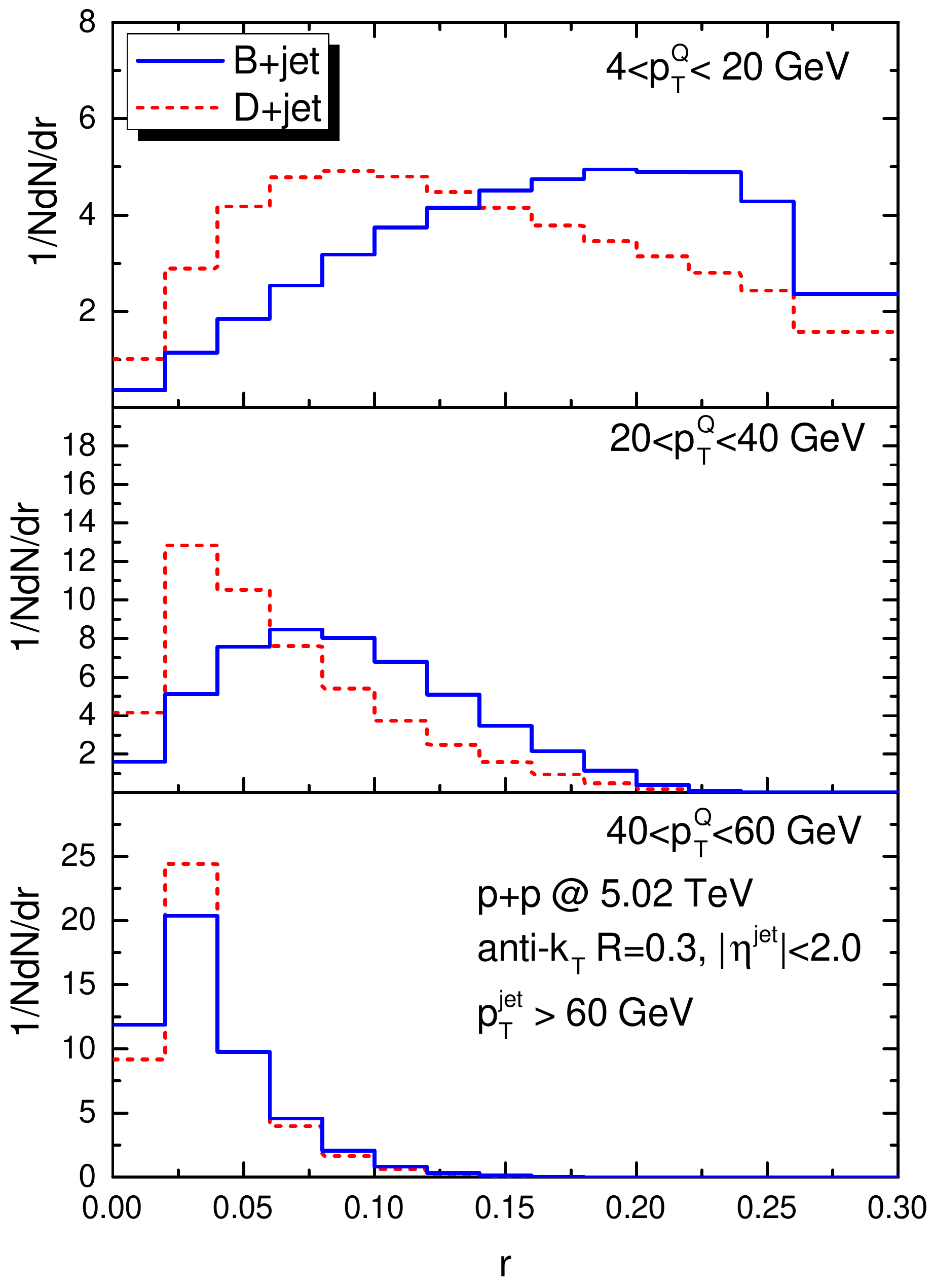}
\vspace*{.1in}
\caption{(Color online) The normalized radial distributions of B meson and D meson in jets as functions of the angular distance to the jet axis in p+p collisions at 5.02 TeV simulated by SHERPA. Three $p_T$ region of heavy quark mesons are plotted as Top, Middle and Bottom panel: $4-20$~GeV, $20-40$~GeV, $40-60$~GeV.}
\label{fig:radia-cb-pp}
\end{center}
\end{figure}

In Fig.~\ref{fig:radia-cb-pp}, we present the calculation results of radial distributions of B mesons in jets in p+p collisions at $\sqrt{s_{NN}}=5.02$~TeV at three transverse momentum regions of B meson ($4<p_T^Q<20$~GeV, $20<p_T^Q<40$~GeV, $40<p_T^Q<60$~GeV) compared with the case of D mesons in jets. $r=\sqrt{(\Delta\phi_{JQ})^2+(\Delta\eta_{JQ})^2}$ is defined by the relative azimuthal angle $\Delta\phi_{JQ}$ and relative pseudorapidity $\Delta\eta_{JQ}$ between the heavy quark meson and the jet axis, please note that the superscript and subscript $Q$ denotes heavy quark meson. All the selected B meson+jet and D meson+jet are reconstructed by anti-$k_T$ algorithm with $R=0.3$ within $|\eta^{\rm jet}|<1.6$, and are required to have $p_T^{\rm jet}>60$~GeV.

When we compare the radial distribution of the same heavy flavor in jets at different $p_T^Q$ regions of the final state heavy flavor meson, we find a narrower distribution of radial profile with the increasing of the $p_T^Q$ both for B mesons and D mesons in jets. We also find the discrepancy of the radial profile distributions of B mesons and D mesons in jets at the same $p_T^Q$ region, especially at $4<p_T^Q<20$~GeV, the B mesons distributed at larger $r$ peaked around $r=0.20$ and D mesons distributed at smaller $r$ peaked around $r=0.07$ shown as the top plots of Fig.~\ref{fig:radia-cb-pp}.
With the enhancement of the $p_T^Q$ region, this kind of discrepancy tends to disappear demonstrated as the bottom plots of Fig.~\ref{fig:radia-cb-pp}. By simply changing the value of the quark mass in the calculation in p+p, we confirm the distinct radial profile distributions all arise from the difference of heavy quark masses. Obviously the $4<p_T^Q<20$~GeV region of the final state heavy flavor meson provide an unique opportunity to investigate the potential different in-medium modifications of B mesons and D mesons in jets, and we also need to keep in mind the $p_T^Q$ trigger sensitivity of the radial profile distribution of heavy flavors in jets for further discussion.


\section{SHELL model: in-medium jet evolution}
\label{sec:framework}
To implement the in-medium parton evolution for light partons and heavy quarks simultaneously, we use the p+p events generated by SHERPA~\cite{Gleisberg:2008ta} with vacuum parton shower as input, and then investigate the subsequent in-medium jet evolution in the hot and dense QCD matter with the model of Simulating Heavy quark Energy Loss with
Langevin equations (SHELL)~\cite{Dai:2018mhw,Wang:2019xey,Wang:2020bqz,Wang:2020qwe} .

In the SHELL model, the initial spacial distribution of partons are sampled by a Monte Carlo Glauber model~\cite{Miller:2007ri}.
When parton propagates in the QGP, two important energy loss mechanisms are considered: collisional interaction (elastic scattering with the constituents of medium) and radiative interaction (medium-induced gluon radiation in the inelastic scattering).
In the infinity heavy quarks limit ($p \sim T$, $M\gg T$), the propagations of heavy quarks in the QGP are usually well described by the Langevin equations. When it goes to higher $p_T$ region, as the medium-induced gluon radiation become the dominant mechanism for the heavy quark energy loss, the modified Langevin equations~\cite{Cao:2013ita,Dai:2018mhw,Wang:2019xey,Wang:2020bqz,Wang:2020qwe} are viewed as a effective method to take into account the radiative correction, as shown in Eq.~(\ref{eq:lang2}),

\begin{eqnarray}
&&\vec{x}(t+\Delta t)=\vec{x}(t)+\frac{\vec{p}(t)}{E}\Delta t\\
&&\vec{p}(t+\Delta t)=\vec{p}(t)-\Gamma(p)\vec{p} \Delta t+\vec{\xi}(t)-\vec{p}_g \, ,
\label{eq:lang2}
\end{eqnarray}

 where $\Delta t$ is the time step of the in-medium Monte Carlo simulation, and $\Gamma$ the drag coefficient. $\vec{\xi}(t)$ is a white noise representing the random kicks obeying $\left \langle \xi^i(t)\xi^j(t') \right \rangle =\kappa \delta^{ij}\delta(t-t')$, where $\kappa$ is the diffusion coefficient in momentum space. $\Gamma$ and $\kappa$ are usually associated by the fluctuation-dissipation relation $\Gamma=\frac{\kappa}{2ET}=\frac{T}{D_s E}$, where $D_s$, the spatial diffusion coefficient , is viewed as a free parameter estimated in various theories~\cite{Rapp:2018qla}. In this study, we chose $D_s=\frac{4}{2\pi T}$ extracted by the Lattice QCD~\cite{Francis:2015daa,Brambilla:2020siz} as a fixed parameter in our calculations. The last term in Eq.~(\ref{eq:lang2}) represents the momentum recoil due to the medium-induced gluon radiation, which is implemented based on the higher-twist calculations~\cite{Guo:2000nz,Zhang:2003yn,Zhang:2003wk,Majumder:2009ge}:

\begin{eqnarray}
\frac{dN}{dxdk^{2}_{\perp}dt}=\frac{2\alpha_{s}C_sP(x)\hat{q}}{\pi k^{4}_{\perp}}\sin^2(\frac{t-t_i}{2\tau_f})(\frac{k^2_{\perp}}{k^2_{\perp}+x^2M^2})^4
\label{eq:dndxk}
\end{eqnarray}

where $x$ and $k_\perp$ are the energy fraction and transverse momentum carried by the radiated gluon. $C_s$ is the quadratic Casimir in color representation, and $P(x)$ the splitting function in vacuum~\cite{Wang:2009qb}, $\tau_f=2Ex(1-x)/(k^2_\perp+x^2M^2)$ the gluon formation time. $\hat{q} \propto q_0(T/T_0)^3$ is the jet transport parameter~\cite{Chen:2010te},where $T_0$ is the highest temperature in the most central A+A collisions. Note that we use the same value $q_0=1.2$~GeV$^2$/fm, determined by a global extraction of the single hadron production in Pb+Pb collisions at the LHC energy~\cite{Ma:2018swx}, to describe the strength of gluon radiation for all partons, the mass effects of heavy quark are included in the last quadruplicate term in Eq.~(\ref{eq:dndxk}).

In consideration of the possible multiple gluon radiation during a time step of our simulation, we assume that the number of radiated gluon obeys the Possion  distribution:

\begin{eqnarray}
P(n)=\frac{\lambda}{n!} e^{-\lambda}
\label{eq:pn}
\end{eqnarray}

where $P(n)$ denotes the probability of $n$ times of radiative interaction during a short time step $\Delta t$, $\lambda$ is the mean value of $n$ and could be estimated numerically by integrating Eq.~(\ref{eq:dndxk}):

\begin{eqnarray}
\lambda(t,\Delta t)=\Delta t\int dxdk^2_{\perp} \frac{dN}{dxdk^2_{\perp} dt}
\end{eqnarray}

During each time step in our simulation, firstly, the total probability, $P(n\ge 1)=1-\lambda e^{-\lambda}$, is calculated to determine whether the radiation occur. If radiation occurs, the number of radiated gluon could be sampled based on Eq.~(\ref{eq:pn}), and subsequently the four momentum of each gluon can be sampled by Eq.~(\ref{eq:dndxk}) one-by-one. It should be noted that, a lower cutoff $\omega_0=\mu_D=\sqrt{4\pi\alpha_s}T$ has been imposed to avoid the divergence in the spectra at $x\rightarrow 0$, namely only the gluon with energy above this cutoff is allowed to emit. This treatment could mimic the detailed balance between gluon radiation and absorption, and then ensure heavy quarks can achieve their thermal equilibrium, $f_{eq}(p)\propto e^{-E(p)/T}$ after enough long propagating time in the QGP medium. The hydrodynamic background profile of the expanding QCD medium is provided by the smooth (2+1)D viscous hydrodynamic model~\cite{Shen:2014vra}. We assume that partons stop the in-medium propagation when their local temperature is under $T_c=165$~MeV. To take into account the initial-state CNM effects in nucleus-nucleus collisions, the nuclear parton distribution function (nPDF) nNNPDF1.0~\cite{AbdulKhalek:2019mzd} has been used in the calculations. It's found that the CNM effects have little impact on the radial profiles of heavy quarks in jets.

The SHELL model has been applied in the studies of medium modification of $p_T$ imbalance of $b\bar{b}$ dijets~\cite{Dai:2018mhw}, correlations of $Z^0$+b-jet~\cite{Wang:2020qwe}. Recently it has also been successfully employed to calculate the angular correlations of $D^0$+jet in p+p and Pb+Pb collisions at $\sqrt{s_{NN}}=5.02~$TeV~\cite{Wang:2019xey,Wang:2020bqz}, and a decent agreement between the model calculations and the experiment measurement has been observed~\cite{Sirunyan:2019dow}.

\section{Medium modification of radial profile of bottom quarks in jets in Pb+Pb collisions}
\label{sec:radial}
The in-medium parton interactions not only dissipate the jet energy to the the hot and dense QCD matter outside the jet cone subsequently suppress the jet production, but also redistribute the energy-momentum of parton inside the jet cone, then alter the jet radial profile and the jet substructure. The modified radial distribution of low $p_T$ heavy quarks relative to their tagged high $p_T$ jets, which act as a reference, could indirectly reflect the dynamical details of the in-medium parton interactions in the hot and dense nuclear matter. In this section, systematic predictions of the radial profile of B mesons in jets both in p+p and Pb+Pb collisions will be presented, and further the comparison between bottom and charm flavors in jets will also be investigated aiming to figure out the impact of the mass effect to the medium modification pattern of the radial profile.

\begin{figure}[!t]
\begin{center}
\vspace*{-0.1in}
\hspace*{-.1in}
\includegraphics[width=3.4in,height=3.8in,angle=0]{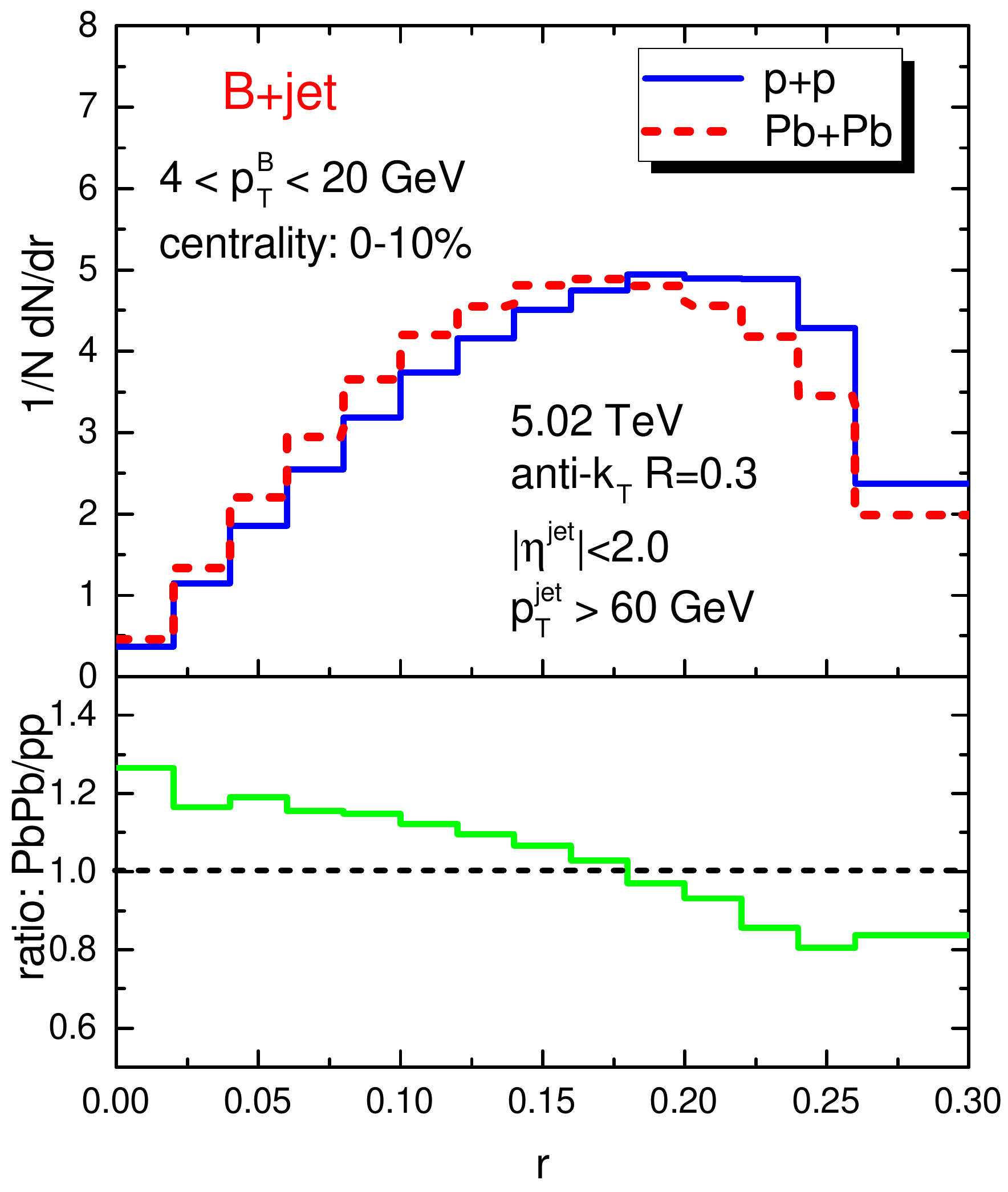}
\vspace*{.1in}
\caption{(Color online) The normalized radial distributions of B meson in jets as a function of the angular distance to the jet axis in p+p and $0-10\%$ Pb+Pb collisions at $\sqrt{s_{NN}}=5.02$ TeV. The ratios of the normalized distribution in Pb+Pb to that in p+p is also plotted in the lower panel.}
\label{fig:radial-cbratio}
\end{center}
\end{figure}

In Fig.~\ref{fig:radial-cbratio}, we predict the B meson radial distribution in jets both in central $0-10\%$ Pb+Pb collisions at $\sqrt{s_\mathrm{NN}} =5.02$~TeV compared to its p+p baseline. Both in p+p and Pb+Pb collisions, the selected jets are required to have $p_T^{\rm jet}>60$~GeV and be tagged by at least one B meson with $4<p_T^Q<20$~GeV, which is the same as our previous study~\cite{Wang:2019xey} and the CMS measurements~\cite{Sirunyan:2019dow} on the case of D mesons in jets. In the top plots in Fig.~\ref{fig:radial-cbratio}, we observe the B meson radial distribution in jets in Pb+Pb shifting towards smaller radii relative to its p+p baseline, therefore find enhancement at smaller radii and suppression at larger radii in the ratio of the normalized radial distribution in Pb+Pb to that in p+p shown in the lower panel of Fig.~\ref{fig:radial-cbratio}. However this kind of modification to a narrower radial profile is in contrast to the towards broader medium modification pattern of D meson radial profile in jets predicted and measured in~\cite{Wang:2019xey,Sirunyan:2019dow}. It is not intuitional to picture the charm quarks shift towards lager radii while bottom quarks shift close to the jet axis due to an identical in-medium parton interaction mechanism without a further investigation. Also an interesting question will be raised: what role is played by the different masses of bottom and charm quarks in jets?

\begin{figure}[!t]
\begin{center}
\vspace*{-0.1in}
\hspace*{-.1in}
\includegraphics[width=3.4in,height=3.1in,angle=0]{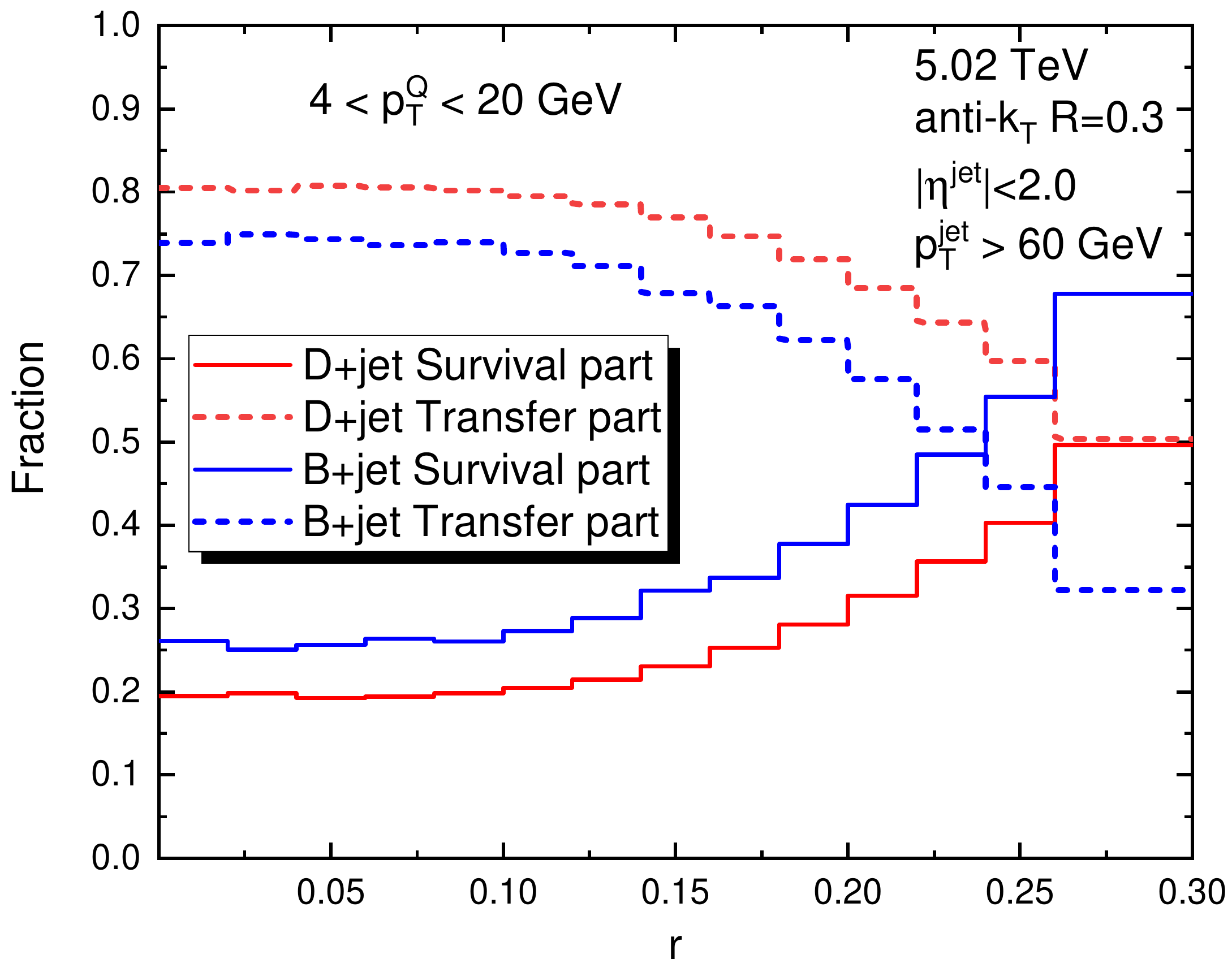}
\vspace*{.1in}
\caption{(Color online) The fractional contribution of the reconstructed event from survival part and transfer part in the B mesons and D mesons radial distributions in jets at $4<p_T^Q<20$~GeV in $0-10\%$ Pb+Pb collisions at $\sqrt{s_{NN}}=5.02$~TeV.}
\label{fig:radial-frac}
\end{center}
\end{figure}

In order to conduct such investigation, it's instructive to firstly clarify the compositions of the jets reconstructed in Pb+Pb collisions. Please note the reconstructed jets in Pb+Pb collisions would also obey the selection of $4<p_T^Q<20$~GeV same as the p+p baseline. Therefore one can easily imagine some of the reconstructed jets in Pb+Pb could be the survived ones whose $p_T^Q$ do not fall below the lower threshold of $4$~GeV due to jet quenching, these survived jets do have their original counterparts in the p+p baseline events before jet quenching with $4<p_T^Q<20$~GeV. However, a lot of the jets reconstructed in Pb+Pb are the ones whose heavy flavor mesons $p_T^Q>20$~GeV before jet quenching. These part of the jets reconstructed in Pb+Pb do not have their counterparts in the p+p baseline events and initially(originally) distributed closer to the jet axis than the case of p+p baseline events simply because of the higher $p_T^Q$ trigger according to the discussion of Fig.~\ref{fig:radia-cb-pp}. To facilitate further discussion, we name these two contribution sources as: Survival part and Transfer part. The separation and the respective investigations of the two contribution sources are not easy for analytical or experimental study, the sources tracking power of our Monte Carlo study can help us to do so and therefore gain more insight of the medium modification mechanism of the heavy flavor radial profiles in jets.

Contribution fraction is always important and essential when talking about the relation of the overall modification and respective modification of each composition. In Fig.~\ref{fig:radial-frac} we plot the contribution fractions of the Survival part and Transfer part in the reconstructed jets in Pb+Pb as a function of $r$ for both D Mesons and B mesons in jets respectively. We find at the $4<p_T^Q<20$~GeV region we investigated in, for both cases of D mesons and B mesons in jets, the Transfer part denoted in short dash line will dominate at smaller $r$ and the domination begin to decrease with the increasing of $r$. We can observe that the charm quark suffers more energy dissipation than bottom quark in such context, since at any $r$ there is always more proportion of the bottom quark survived, even at lager $0.25<r<0.3$, the survival part begins to dominate.

\begin{figure}[!t]
\begin{center}
\vspace*{-0.1in}
\hspace*{-.1in}
\includegraphics[width=3.4in,height=5in,angle=0]{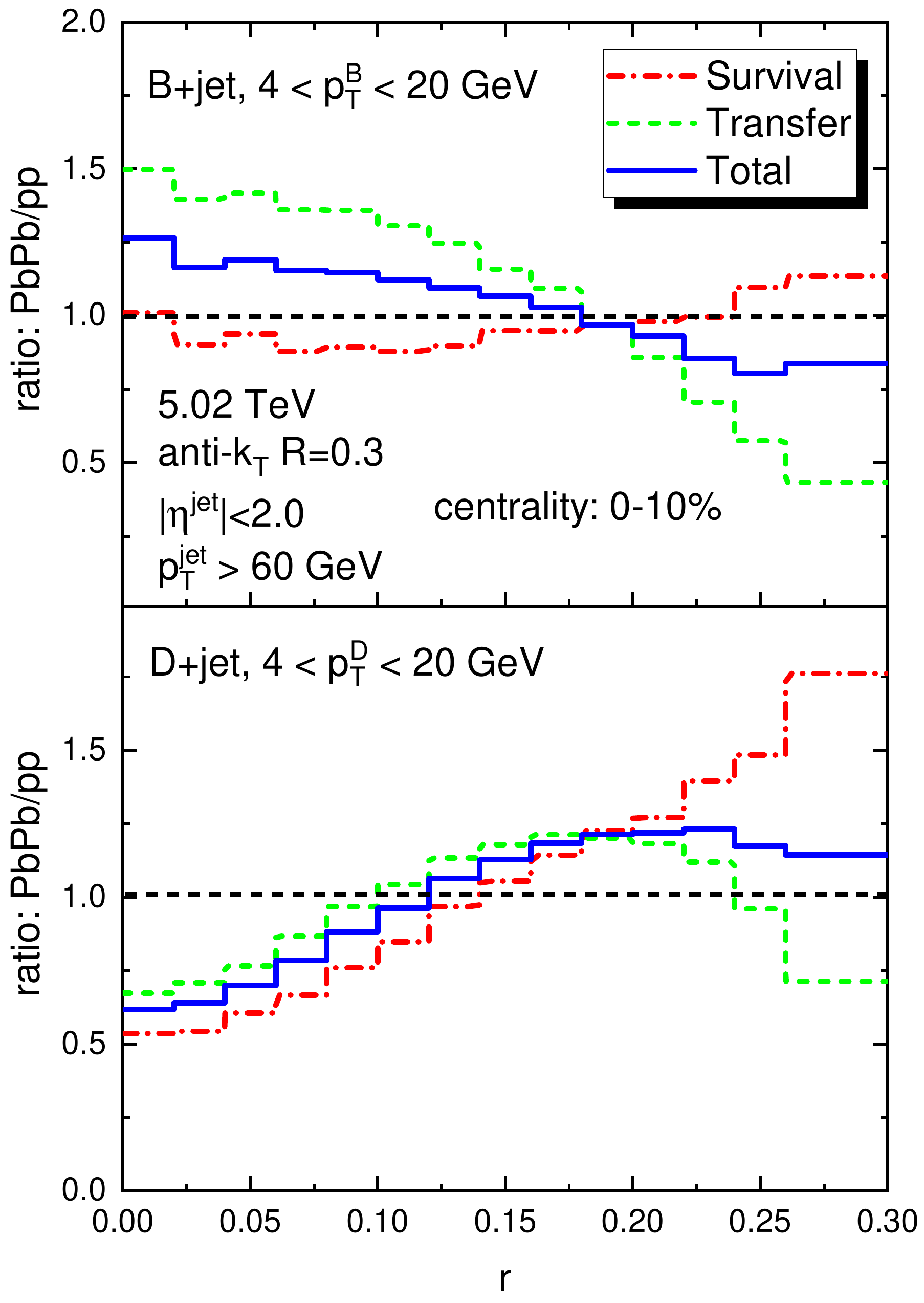}
\vspace*{.1in}
\caption{(Color online) The medium modification patterns on the radial profiles of heavy flavor mesons (upper panel: B meson, lower panel: D meson) in jets from two kinds of contributions: Survival part (red dot-dash line) and Transfer part (green dash line), as well as the total contribution (blue solid line).}
\label{fig:radial-sub}
\end{center}
\end{figure}

With such knowledge of the contribution fractions of the two parts in Pb+Pb events, we plot in Fig.~\ref{fig:radial-sub} the  modification patterns of B mesons radial profile in jets in the top panel and the case for D mesons radial profile in jets in the bottom panel. In the plots, we also include the ratios of the self-normalized radial distribution of the Survival and Transfer contribution parts in Pb+Pb to the normalized p+p baseline denoted as dash dot line and short dash line respectively. Firstly, we do confirm an inverse medium modification pattern of the radial profile of the heavy flavor in jets when comparing the total normalized distribution PbPb/pp ratios as functions of $r$ of B+jet (solid line in the top panel) and D+jet(solid line in the bottom panel). The total normalized distribution PbPb/pp ratio is combinated by the Survival and Transfer contributions taking into account their contribution fractions. Since at smaller $r$ region, Transfer part dominate the contribution fraction, the modification pattern of the Transfer part will be dominant, but at larger $r$ region the contribution fraction will interplay with the respective ratios of the two contribution parts to determine the total ratio.

When we compare the modification patterns of B mesons and D mesons radial profile in jets in Survival and Transfer contributions respectively, we are getting closer to reveal the nature of such inverse medium modification patterns. The comparison of the Survival modification pattern (in dash dot line) of B mesons (top) and D mesons (bottom) radial profile in jets indicate: a same broader modifications of the radial distribution are shown from both the cases of B mesons and D mesons in jets due to the possible diffusion effects which is previous investigated in Ref.~\cite{Wang:2020bqz}. However we observe a relatively smaller diffusion effect in the B mesons radial profile in jets than D mesons. For the case of Transfer contribution, we find a narrower modification pattern of B mesons radial profile in jets but a broader modification pattern of D mesons in jets, the two opposite modification directions are mainly due to the competition of two effects which lead to opposite consequences. One is that the Transfer part originally comes from higher $p_T^Q$ trigger and naturally distributed at smaller $r$, when the ratio to the p+p baseline is taken, distribution shifting towards narrower direction will observed. Meanwhile the other fact is that the in-medium modification of such heavy flavor will lead to a broader direction of distribution shifting due the diffusion effect. These two effects will offset with each other. The first effect reveals the energy dissipation nature and the latter shows the diffusion feature that always lead to spreading away from the jet axis.

From the discussion of Fig.~\ref{fig:radia-cb-pp}, the $p_T^Q$ sensitivity of the initial radial distribution of B mesons in jets is larger than D mesons in jets and it will lead to a larger narrower shifting towards smaller $r$ of the radial distribution of B mesons in jets than D mesons in jets. Consequently, a larger narrower shifting towards smaller $r$  due to larger $p_T^Q$ sensitivity of the initial radial distribution compared to charm quarks in jets and a smaller broader shifting towards smaller $r$ due to its weaker diffusion effect compared to charm quarks in jets will lead to an overall narrower shifting which is exactly opposite with the case of D mesons in jets.

\begin{figure}[!t]
\begin{center}
\vspace*{-0.1in}
\hspace*{-.1in}
\includegraphics[width=3.4in,height=3.8in,angle=0]{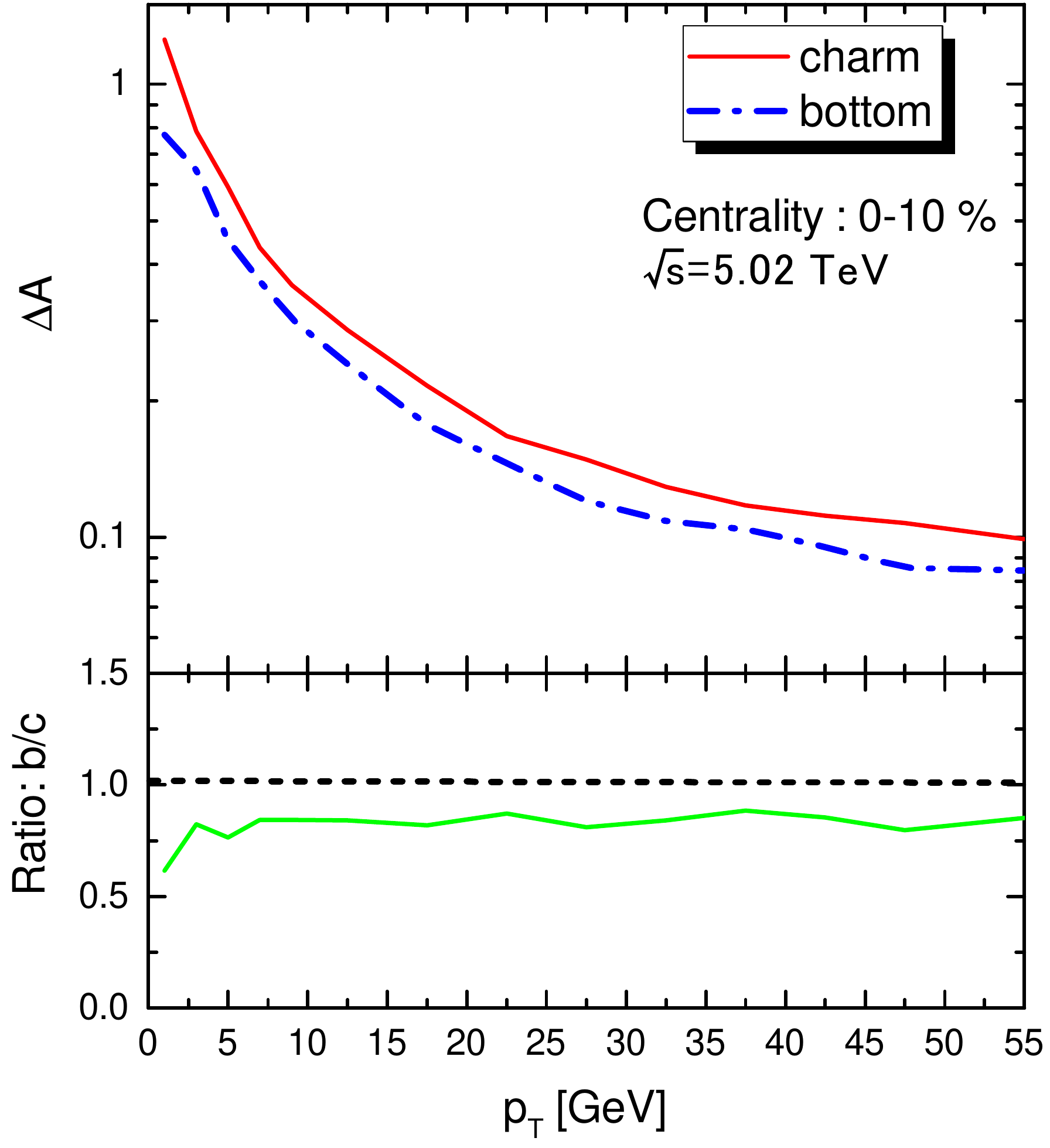}
\vspace*{.1in}
\caption{(Color online)  Angular deviation ($\Delta A$) of charm and bottom quarks relative to their initial moving direction in the medium as a function of transverse momentum, and the ratio of charm to bottom was shown at the lower panel.}
\label{fig:drdpt-cb}
\end{center}
\end{figure}

To further demonstrate the angular diffusion effect of bottom quarks is much weaker as compared to that of charm quarks in the medium, we firstly consider the impact of the interaction strength between charm and bottom quarks in the QGP. We define $$\Delta A=\sqrt{(\eta_Q-\eta_Q^0)^2+(\phi_Q-\phi_Q^0)^2}$$ to quantify the angular deviation of the heavy quarks from their original moving directions during the in-medium propagation, where $\eta_Q^0$ and $\phi_Q^0$ are the initial pseudorapidity and azimuthal angle of heavy quarks before entering the QGP medium. The angular deviation $\Delta A$ represents directly the diffusion effect of the heavy quarks in the $\eta-\phi$ plane due to collisional and radiative energy loss process in the QGP medium. In the top panel of Fig.~\ref{fig:drdpt-cb}, we calculate the angular deviation of charm and bottom quarks after the in medium modification as a function of their initial transverse momentum before energy loss in central Pb+Pb collision at $\sqrt{s_{NN}}=$5.02 TeV. It turns out that diffusion strength decreases with the initial transverse momentum of heavy quarks, and the $\Delta A$ of bottom quarks is smaller than that of charm quarks by nearly $20-30\%$ shown in the bottom panel of Fig.~\ref{fig:drdpt-cb}.

\begin{figure}[!t]
\begin{center}
\vspace*{-0.1in}
\includegraphics[width=3.4in,height=3in,angle=0]{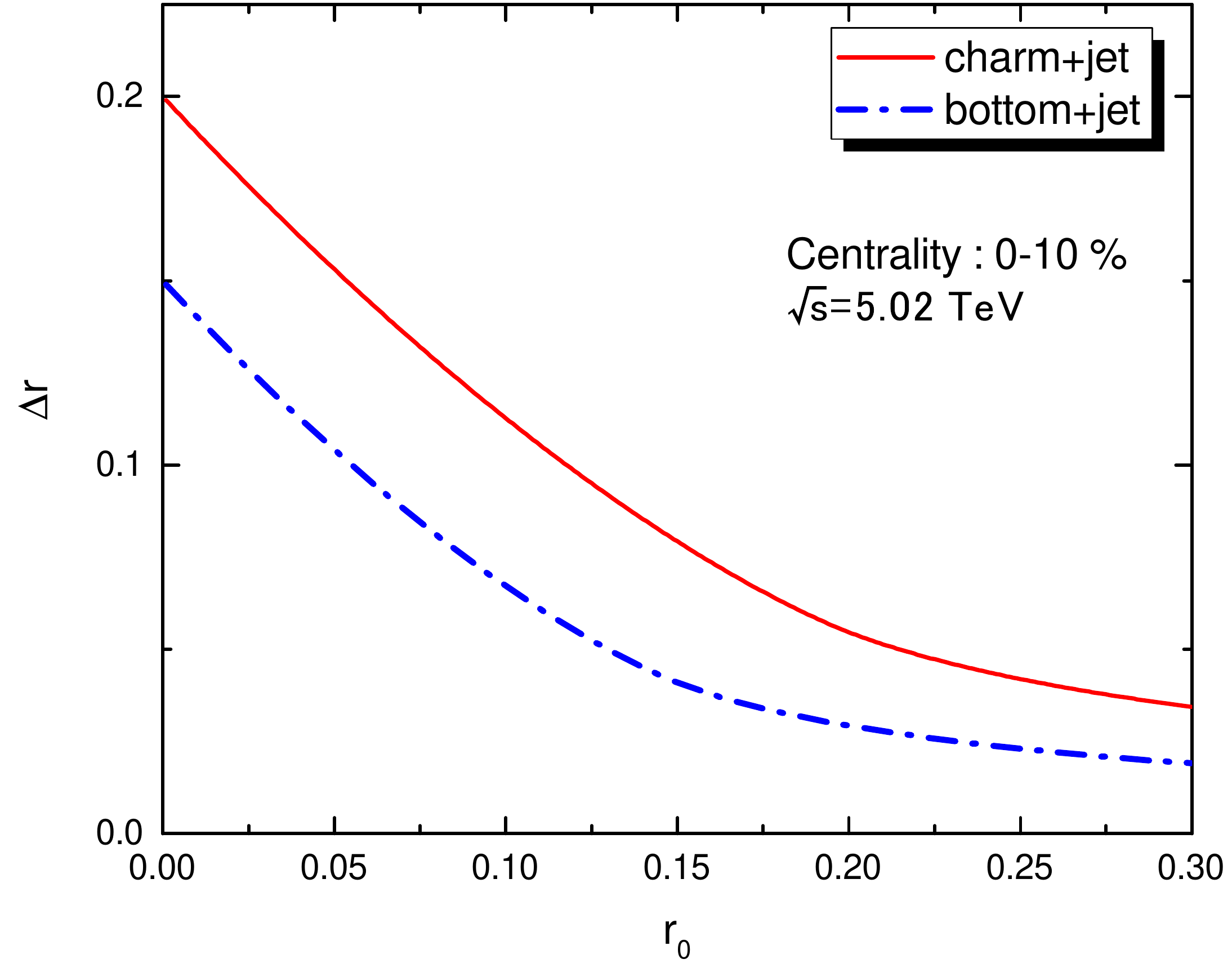}
\vspace*{.1in}
\caption{(Color online) Angular shift ($\Delta r=r-r_0$) of charm and bottom relative to the jet axis in the medium as a function of the initial angular distance to the jet axis ($r_0$).}
\label{fig:drr-cb}
\end{center}
\end{figure}

To explore further the distinction between the medium modifications of bottom quarks and charm quarks radial distributions in jets, we now consider the impact of the initial angular distance between the heavy quarks and the jet axis since they are quite different for the case of D mesons and B mesons in jets in the same $p_T^Q$ region in p+p. We plot in Fig.~\ref{fig:drr-cb} the final observed angular shift $\Delta r=r-r_0$ of both D mesons and B mesons in jets in $0-10\%$ Pb+Pb collisions at $\sqrt{s_{NN}}=5.02$~TeV as functions of their initial angular distance with the jet axis $r_0$. We find the angular shift of charm quark is stronger than that of bottom quark even at the same $r_0$, also the closer the heavy flavor quark is initially distributed away from the jet axis, the larger final observed angular shift $\Delta r$ will be. From the knowledge of Fig.~\ref{fig:radia-cb-pp}, at the same $p_T^Q$ trigger, the charm quark is always distributed closer with the jet axis than bottom quark, therefore it will eventually further enhance the difference of angular shift $\Delta r$ between D mesons and B mesons in jets observed at final-state.

In the calculations $\hat{q}=q_0 (T/T_0)^3$ is assumed, and we note that some possible non-perturbative non-conformal variations of $\hat{q}(T,E)$ near the critical temperature $T_{c}$ were suggested in recent studies~\cite{Xu:2015bbz,Gyulassy:2020jlb}. It will be interesting to see how this non-conformal variation of $\hat{q}$ will influence the jet radial distribution in future's studies, which may deepen our understanding of the QCD phase transition  in the cross-over temperature range.

\section{Summary and conclusions}
\label{sec:summary}

In this work, we present a theoretical investigation of medium modifications of radial distributions of bottom quarks in jets in Pb+Pb collisions relative to that in p+p. We carry out the numerical calculations within a Monte Carlo simulation framework which utilizes the NLO+PS event generator SHERPA as input, and the SHELL model to take into account the in-medium jet energy loss for both heavy and light partons. In p+p collisions, we find that at lower $p_T^Q$ the radial profiles of heavy flavors in jets are sensitive to the heavy quark mass: bottom quarks trend to distribute at the region more far away from the jet axis due to their larger mass compared to charm quarks.

Furthermore, in order to investigate the mass effect of heavy quarks reflected in the in-medium diffusion effect, we estimate the medium modification of the radial distribution of bottom quarks in jets, and we find inverse modification pattern compared to that of charm quarks: jet quenching effects will narrow the jet radial profile of bottom quarks while broaden that of charm quarks. By analyzing the event composition in Pb+Pb collisions, we find the main contribution of the event selected with $4<p_T^Q<20$~GeV in Pb+Pb collisions is dissipated from higher $p_T^Q$ region due to energy loss, namely the contribution of Transfer part. We reveal that both for the cases of  B mesons and D mesons in jets in A+A collisions, the Transfer part naturally has a narrower initial distribution and consequently leads to a narrower modification pattern of radial profile; however the diffusion nature of the heavy flavor in-medium interaction will give raise to a broader modification pattern of radial profile. These two effects consequently compete and offset with each other. In the investigated $\rm p_T^Q$ region, we demonstrate that the b quarks in jets benefit more from the dissipation contribution from higher $\rm p_T^Q$ region and suffers less diffusion effect compared to that of c quarks in jets. Hopeful the sharp contrast of medium modification patterns between radial profiles of charm and bottom mesons in jet at $4<p_T^{Q}<20$~GeV predicted in this work could be tested against the future experimental measurements at the LHC.

The flavor dependence of jet quenching has been studied extensively and for so long~\cite{ vanHees:2007me,CaronHuot:2008uh,vanHees:2005wb,Djordjevic:2015hra,He:2014cla,Chien:2015vja,Kang:2016ofv,Cao:2013ita,Alberico:2013bza,Xu:2015bbz,Cao:2016gvr,
Das:2016cwd,Ke:2018tsh,Xing:2019xae,Li:2020kax,Li:2020umn,Altenkort:2020fgs,Jamal:2020fxo,He:2019vgs}, however, the past studies of this kind mainly focus on the  magnitude difference of heavy quark energy loss due to ``dead-cone'' effect. In this paper, we demonstrate that the medium modifications of radial profiles of a heavy quark in jets are determined not only by how much the heavy quark loses its energy in the QGP,  but also more interesting by the initial radial distributions of heavy quarks in jets, which exhibit quite distinct behaviors at  different $p_T^Q$ kinematic cut as shown in Fig.~\ref{fig:radia-cb-pp}.  Therefore we observe the different initial distributions can ultimately lead to inverse medium modification patterns of B+jet and D+jet at $4<p_T^Q<20$ GeV in Pb+Pb collisions. It reveals a fact that for some jet observables, energy loss mechanism may not be the only factor to determine the final-state modification pattern in nucleus-nucleus collisions, and the initial differential distributions in p+p collisions may also play a very important role, which will then lead to much richer phenomena to be observed in the flavor dependence of jet quenching in heavy-ion collisions.

{\it Acknowledgments:}
This research is supported by the Guangdong Major Project of Basic and Applied Basic Research No. 2020B0301030008, the Science and Technology Program of Guangzhou No. 2019050001 and the NSFC of China with Project Nos. 11935007, 11805167.

\end{document}